# High-throughput, high-resolution registration-free generative adversarial network microscopy

Hao Zhang[1+], Chunyu Fang[1+], Xinlin Xie[1], Yicong Yang[1], Di Jin[3*] and Peng Fei[1, 2, 4*]

[1] *School of Optical and Electronic Information, Huazhong University of Science and Technology, Wuhan, 430074, China.*

[2] *Britton Chance Center for Biomedical Photonics, Wuhan National Laboratory for Optoelectronics, Huazhong University of Science and Technology, Wuhan, 430074, China.*

[3] *Computer Science and Artificial Intelligence Laboratory, Massachusetts Institute of Technology, Cambridge, MA, 02139, U.S.A.*

[4] *Shenzhen Huazhong University of Science and Technology Research Institute, Shenzhen, 518000, China.*

*Correspondence: jindi15@mit.edu and feipeng@hust.edu.cn

[+] equal contributing author

**Abstract**: We combine generative adversarial network (GAN) with light microscopy to achieve deep learning super-resolution under a large field of view (FOV). By appropriately adopting prior microscopy data in an adversarial training, the neural network can recover a high-resolution, accurate image of new specimen from its single low-resolution measurement. Its capacity has been broadly demonstrated via imaging various types of samples, such as USAF resolution target, human pathological slides, fluorescence-labelled fibroblast cells, and deep tissues in transgenic mouse brain, by both wide-field and light-sheet microscopes. The gigapixel, multi-color reconstruction of these samples verifies a successful GAN-based single image super-resolution procedure. We also propose an image degrading model to generate low resolution images for training, making our approach free from the complex image registration during training dataset preparation. After a well-trained network being created, this deep learning-based imaging approach is capable of recovering a large FOV (~95 mm$^2$), high-resolution (~1.7 μm) image at high speed (within 1 second), while not necessarily introducing any changes to the setup of existing microscopes.

1. Introduction

The imaging throughput of a conventional optical microscope is typically limited to megapixels, regardless of the magnification and numerical aperture used[1, 2]. As a result, compromise often exists between achieving a high resolution and maintaining a large field-of-view (FOV). However, nowadays high-resolution mapping of entire large specimens is increasingly desirable for life science applications such as tissue pathology, hematology, digital histology and neuron science[3, 4]. In order to precisely interpret cellular events throughout entire samples, global structures and local details spanning from micro- to meso-scale need to be continuously measured and quantitatively analyzed at the same time[5]. Development of sophisticated mechanical scanning microscope is a commonly-used way to address this challenge, artificially increasing the throughput of the microscope by stitching multiple high-resolution tiles into a panoramic image[6].

Besides this mechanical approach that requires precise control over actuation and optical alignment, recent super resolution (SR) techniques present a computational way to increase the space-bandwidth product of a microscope platform[1, 7-19] For instance, pixel super resolution (PSR) represents a class of spatial domain techniques that can fuse multiple large FOV, low resolution measurements with sub-pixel shifts into a high resolution image[17, 18]. On the other hand, several frequency domain methods, e.g., Fourier ptychographic microscopy (FPM)[1], synthetic aperture microscopy[7-10] and structured-illumination microscopy[20, 21], produce a resolution-enhanced image by stitching together a number of variably illuminated, low-resolution images in Fourier domain. Despite offering unique imaging capabilities with scalable SBP, these methods, however, all require special hardware setup and complex computation on multiple frames. Nevertheless, another type of technique, named single image super resolution (SISR), has been widely applied in microscopy without these constraints. It aims at the reconstruction of a high-resolution (HR) images with rich details from single low-resolution (LR) image. For this technique, the conventional widely used method is the example-based approach[22, 23], which works by replacing the LR information with the HR patches searched out in the example dictionary. Although SISR greatly simplifies the imaging hardware architecture and the computation complexity, the quality of reconstructed images remains suboptimal as compared to the multi-frame methods. The recent advent of deep learning neural network is providing another way to realize more efficient SISR. Artificial neural network based super-resolution has been demonstrated on bright-field microscopy[24, 25] as well as fluorescence microscopy[26-29]. Most recent model that utilizes the generative adversarial network (GAN) for better visual details enhancement, has reached remarkable resolution enhancement[26, 29]. However these methods require an extra image registration between high-resolution and low-resolution training pairs captured under different magnifications. Considering a pixel-wise error function is the most common practice in super resolution, the accuracy of registration could affect the performance of the neural network.

Here we present a deep learning-based super resolution approach that is free from registration during training process, meanwhile capable of providing significant resolution enhancement for conventional microscopy, without the need of acquiring a plurality of frames or retrofitting existing optical systems[30]. This imaging method uses datasets that consist of high-resolution measurements and their low-resolution simulations to train a GAN model. We carefully model the image degradation of the microscope system to generate low-resolution trial images from measured high-resolution source images, thereby eliminating the need of complicated alignment between the high- and low-resolution pairs. As long as the network training with prior data is accomplished, the network is capable of using single low-resolution measurement of a new specimen to recover its high-resolution, large FOV image. We demonstrate the efficiency of this registration-free GAN microscopy (RFGANM) method with bright-field image of USAF resolution target, color image of whole pathological slides, dual-channel fluorescence image of fibroblast cells, and light-sheet fluorescence image of a whole mouse brain, verifying that it's widely applicable to various microscopy data. By taking a few example images as the references and applying a GAN deep-learning procedure, we can transform a conventional optical microscope into a high-resolution (~1.7 μm), wide-FOV (~95 $mm^2$) microscope with a final effective SBP of 0.13 gigapixels. Furthermore, the reconstruction procedure is independent from the GPU-based training, thus can be performed on a normal computer without GPU installed. This underlying advantage renders RFGANM a robust platform that allows multiple applications to be followed once after a well-trained SR artificial intelligence based system is established. In the following, we will briefly describe the RFGANM operation and experimental set-up, discuss how to apply the network training and inference process, and discuss its imaging applications in a variety of biomedical samples.

## 2. Results

*2.1. Deep learning based image super resolution reconstruction.*

A classic GAN model [31] that consists of a generator and a discriminator, is used to "learn" the various types of microscopy data from scratch. Figure 1 illustrates the network training and inference process. We establish its capability of mapping from a LR image to a HR reconstruction as shown in Fig. 1a. Firstly multiple HR images of the example specimen are captured under high-magnification objective (Fig. 1a, step 1), then through accurately modeling the transfer function of the microscope system, we can obtain the down-sampled, blurred images of the example specimen directly via simulation (Fig. 1a, step 2). Based on its currently-learned parameters, the generator creates resolution-enhanced reconstructions of LR simulations in each training iteration (Fig. 1a, step 3). The differences between the generator outputs and the realistic HR images are calculated using the mean squared error (MSE), denoted as the content loss function of the generator (Fig. 1a, step 4). Besides the generator, GAN includes an additional discriminator that aims to evaluate the reliability of the generator. This discriminator makes a judgement on whether an image is a reconstruction by the generator or a realistic high-resolution measurement, after they are randomly input (Fig. 1a, step 5). An adversarial loss is created to estimate the accuracy of the discriminator's judgement. It iteratively optimizes the discriminator, aiming at an enhanced capability on making correct decision. Also, the adversarial loss together with the content loss are used to optimize the generator, pushing it towards the direction that generates more perceptually realistic outputs which can further fool the discriminator. This adversarial training process thereby promotes the accuracy of both the generator and the discriminator. The training process can be terminated when the generator produces results that the discriminator can hardly tell from the realistic HR images. Then in the inference phase, a LR measurement of sample, which is excluded from the training dataset, is divided into several patches and fed into the well-trained generator (Fig. 1b, step 6). The generator is capable of recovering high frequency information for each patch, based on the prior GAN training. These quality-improved patches are finally stitched into one gigapixel image of the sample that encompasses high-resolution details as well as large FOV (Fig. 1b, step 7). The aforementioned image reconstruction process is illustrated in Figure 1b, and the overall implementation process of our approach is illustrated in Fig. 5a. It is noteworthy that usually the GAN training is required only once, and then applicable to the recovery of multiple samples with similar type of signals.

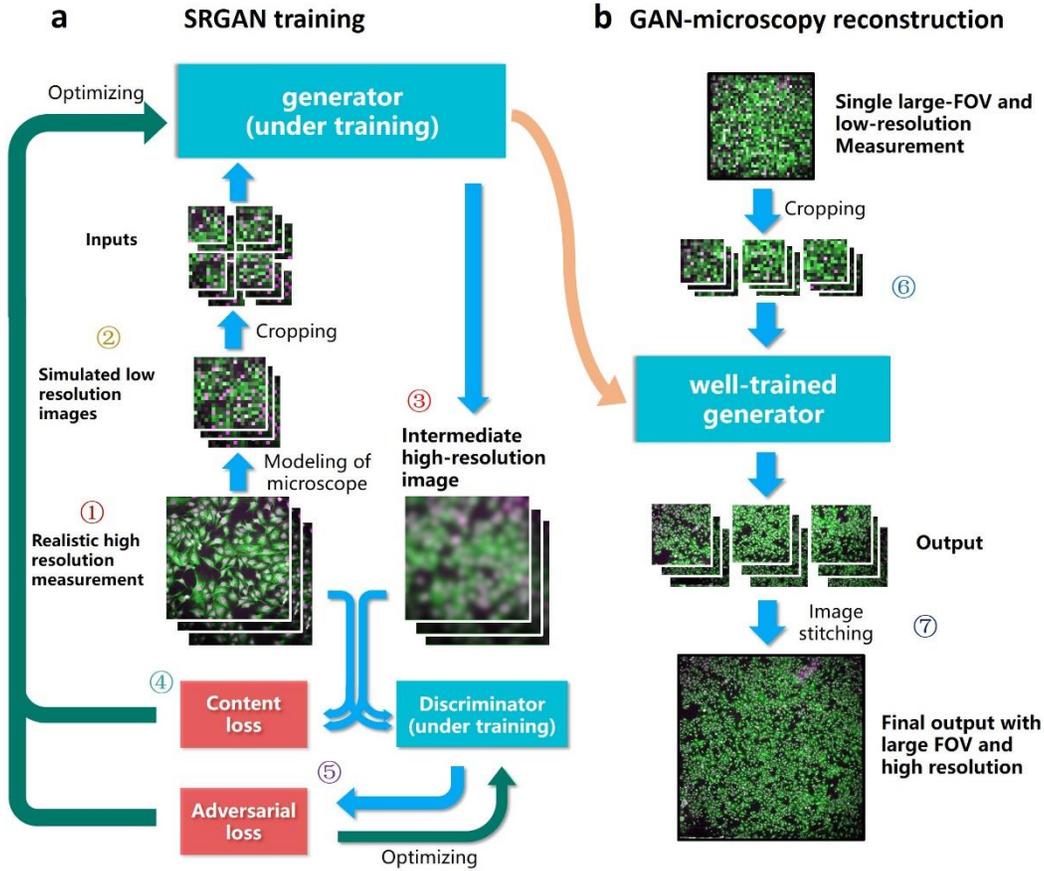

**Figure 1. Principle of RFGANM procedure.** A modified GAN is used to iteratively learn the microscopy data till the goal of high-quality output is reached. **a**, The training process of the RFGAN. **b**, The RFGANM reconstruction procedure, using a large-FOV, low-resolution measurement of new sample.

## 2.2. Image degrading model.

It is widely accepted that the performance of a neural network relies heavily on the training dataset, where there are LR images as inputs and HR images as targets for super resolution task. These LR and HR image pairs for training can be obtained in two ways. Most intuitively, both LR and HR images are experimentally captured with microscope. However, since the LR and HR image pairs are taken under different magnifications, image cutting and registration techniques must be used to match the FOV and remove the unavoidable distortion. Therefore, the performance of image registration is the key to the quality of training data, which is mainly based on feature detection and matching. Unfortunately, in cases of cell, tissue and brain imaging, a great deal of feature details is lost in LR images compared with the corresponding HR images due to the down-sampling process, leading to a high failure rate of image registration. Even though we have used a decent and standard image registration procedure, the mismatch between LR and HR images happens a lot, which significantly deteriorates the quality of training dataset.

Instead of capturing LR and HR images under different magnifications and then aligning them, we can apply an image degrading model to the captured HR images to generate the simulated LR images. In a nutshell, the LR images for training are directly down-sampled from the HR images, so we can guarantee that the two images share the same FOV. To make sure that our model trained on the simulated LR images can still well super-resolve the experimentally captured LR images, the image degrading model we used should be able to produce a simulated LR image as close to the captured LR counterpart as possible. The degrading process of conventional microscopy system can be described as

$$I_m = D(K * I) + N \qquad (1)$$

where $I$ is the continuous real intensity distribution of the sample to be imaged; $K$ is the point spread function of the optical system, represented as a Gaussian convolution kernel; operator $*$ is the convolution between $I$ and $K$; $D$ acting on the convolution results denotes the discretization by the camera sensor; $N$ denotes the additive Gaussian white noise, mainly contributed by the statistic thermal noise of the CCD/CMOS sensor; and $I_m$ is the digital measurement we obtained, which is the discrete, decimated approximation of $I$.

In practice, the real distribution $I$ is approximated by a high-resolution measurement obtained under high-magnification optics. Discretization operation $D$ is thereby a down-sampling on the digital image. There are two parameters to be optimized: size of the Gaussian kernel in the convolution step and variance of the noise distribution. This trail-and-error procedure is visualized in Figure 2. Taking the degradation of a x20 cell image as an example, we de-noise the x4 measurement, comparing it with the blurred, down-sampled x20 measurement to figure out the proper sigma value of the Gaussian kernel. After an optimal sigma value is found, we fix it and add noise with different variance, comparing the degraded result with the original x4 measurement to identify the proper variance. After the degrading model being established, we verify it multiple times. The difference between the simulation and the measurement is directly computed via the pixel-wise subtraction, as shown in Figure 2g. Except for slight mismatch of the cell shape caused by the imaging aberration, the differences are nearly Gaussian white noise, demonstrating the successful application of the degrading model.

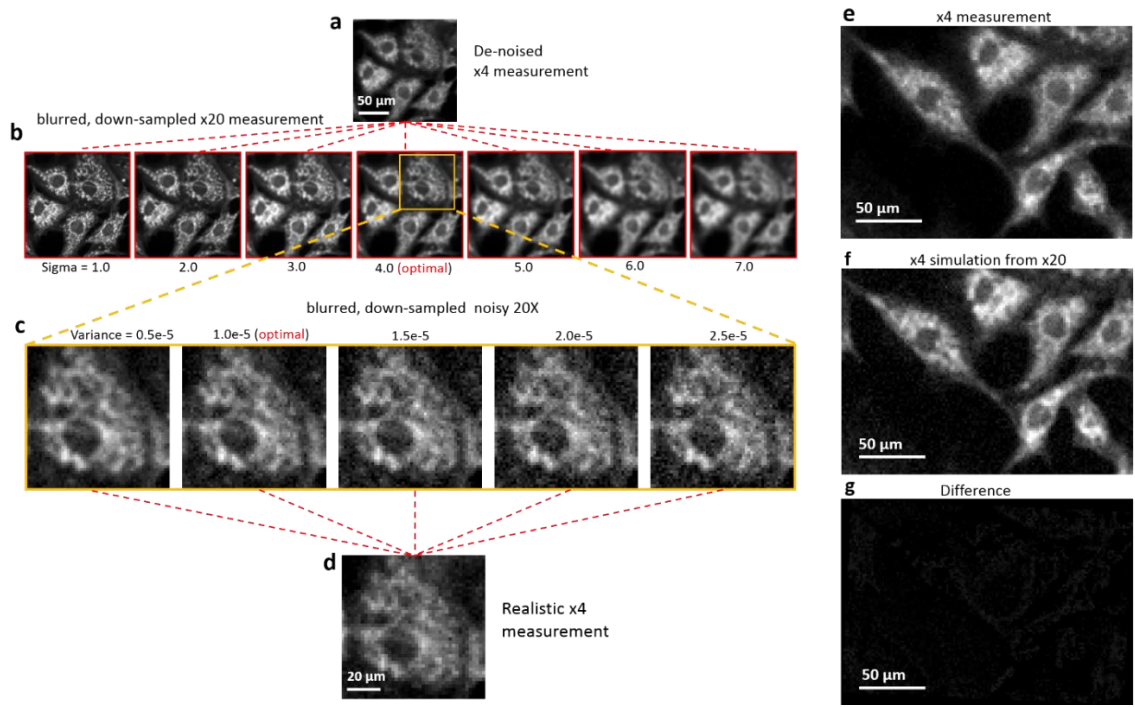

**Figure 2. Generation of simulated LR inputs via a degradation model. a,** De-noised x4 LR measurement used for finding the optimized blurring kernel of microscope. **b,** The blurred and down-sampled images from x20 measurement, by different sizes of blurring kernel (Sigma value). **c,** Blurred image with optimal blurring is processed with different additive noises and compared to the realistic x4 measurement. The best matched level of noise is found as a result. **e to g,** a x4 measurement is subtracted by the optimized x4 simulation to verify the efficacy of the model.

After generating the simulated LR images using the captured HR measurements, we cut the images into small pieces to compose the training dataset, of which the LR pieces are 96×96 pixels and HR pieces are 384×384 pixels. Each LR piece has a corresponding HR one with the same FOV. By cutting the images, we can increase the number of training data a lot, and smaller image size is more compatible to the limited GPU memory. We also adopt several widely-used data augmentation methods, such as translation, rotation and flipping, to further expand the training dataset.

Implementing appropriate GAN-training with simulations generated by an accurate degrading model is crucial for the high quality SR outputs. The accurate and not accurate LR simulations used as the training data result in very different network outputs, as shown in Figure 3. Taking the same LR measurement as the input, network trained with finely-tuned simulations generates far better results than the one trained with inaccurate simulations.

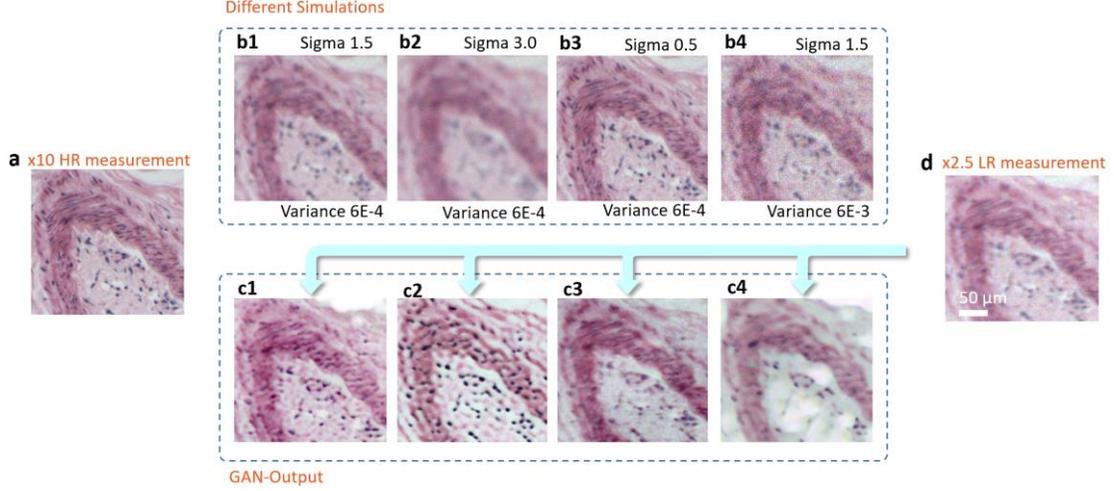

**Figure 3. Network outputs with different degrading parameters. a**, x10 measurement of a human stomach tissue slice. **b1-b4**, LR simulations degraded from a, with different parameters applied. Sigma is the standard deviation of the Gaussian blur kernel and variance denotes the variance of the white noise distribution. **c1-c4,** After the network being trained by different simulation data, it correspondingly generates different reconstruction results that are recovered from the same LR measurement (**d).**

*2.3. Neural net architecture and loss functions.*

The model architecture of the GAN is illustrated in Figure 4. In the generator, the input image is first processed by one convolution layer of 3 x 3 kernel followed by ReLU activation and then goes through 16 residual blocks with identical layout. In each residual block, there are two convolutional layers with small 3 x 3 kernels and 64 feature maps followed by batch-normalization (BN) layers and ReLU as the activation function, and the input and output are element-wise added via shortcut connections. To enhance the residual learning, the output feature map of the first convolution layer and that of the final residual block is also element-wise added. We increase the resolution of image using two layers of sub-pixel convolution suggested by Shi et al.[32] .

To optimize the generator, we designed the perceptual loss function that is the weighted sum of the MSE loss $\ell^G_{MSE}$, the feature reconstruction loss $\ell^G_{feat}$ and the adversarial loss $\ell^G_{adv}$ as proposed by Ledig et al.[33]:

$$\ell^G = \frac{1}{N}\sum_{n=1}^{N}\left(\ell^G_{MSE}\left(G_{\theta_G}(\Delta(I_n^{HR})), I_n^{HR}\right) + 20^{-6}\ell^G_{feat}\left(G_{\theta_G}(\Delta(I_n^{HR})), I_n^{HR}\right) \right.$$
$$\left. + 10^{-3}\ell^G_{adv}\left(G_{\theta_G}(\Delta(I_n^{HR}))\right)\right)$$

(2)

where $G_{\theta_G}$ is the generator parameterized by $\theta_G$, $I^{HR}$ is the high resolution measurement and $\Delta$ is the degeneracy course by our degrading model. In this equation, the MSE loss measures the pixel-wise difference between the output super-resolved image and the target high-resolution image, calculated as:

$$\ell_{MSE}^G(G_{\theta_G}(\Delta(I_n^{HR})), I^{HR}) = \frac{1}{r^2WH}\sum_{x=1}^{rW}\sum_{y=1}^{rH}(I_{x,y}^{HR} - G_{\theta_G}(\Delta(I_n^{HR}))_{x,y})^2 \tag{3}$$

Besides using the GAN framework to encourage perceptual similarity, we further used the special feature reconstruction loss function proposed by Johnson et al.[34]. Let $\phi_j(x) \in R^{W_j*H_j*C_j}$ be the activations of the *j*th convolution layer of the VGG19 network described in Simonyan and Zisserman[35] when processing the image. Then the feature reconstruction loss is defined by the Euclidean distance between the feature representations of the reconstructed image $G_{\theta_G}(\Delta(I_n^{HR}))$ and the reference image $I^{HR}$:

$$\ell_{feat/j}^G(G_{\theta_G}(\Delta(I_n^{HR})), I^{HR}) = \frac{1}{W_jH_j}\sum_{x=1}^{W_j}\sum_{y=1}^{H_j}(\phi_j(I^{HR})_{x,y} - \phi_j(G_{\theta_G}(\Delta(I_n^{HR})))_{x,y})^2 \tag{4}$$

where *j* in our experiments was set to 12.

In addition to the losses described so far, we also need to add the adversarial component of our GAN for the generative side to the perceptual loss. It is defined based on the probabilities of the discriminator over the reconstructed samples as:

$$\ell_{adv}^G\left(G_{\theta_G}(\Delta(I_n^{HR}))\right) = -\log D_{\theta_D}(G_{\theta_G}(\Delta(I_n^{HR}))) \tag{5}$$

where $D_{\theta_D}$ is the discriminator parameterized by $\theta_D$. For better gradient computation stability[31], we minimize $-\log D_{\theta_D}(G_{\theta_G}(\Delta(I_n^{HR})))$ instead of $\log(1 - D_{\theta_D}(G_{\theta_G}(\Delta(I_n^{HR}))))$.

As for the discriminator, It first contains 8 convolutional layers with 4 x 4 kernels followed by BN layers and LeakyReLU (α=0.2) activation (except that the first convolutional layer does not come with BN). Through these 8 layers, the feature map dimension first increases gradually by a factor of 2 from 64 to 2048, then decreases by the same factor to 512. Strided convolutions with stride of 2 are used to reduce the image resolution each time the number of features is doubled. Afterwards, the network is followed by a residual block that contains three convolutional layers followed by BN and LeakyReLU activation. Finally, the resulted 512 feature maps are flattened and connected by one dense layer and a sigmoid activation function to obtain the final probability over whether the input image is natural or not. The network is trained by minimizing the following loss function:

$$\ell^D = \frac{1}{N}\sum_{n=1}^{N}(-\log D_{\theta_D}(I_n^{HR}) - \log(1 - D_{\theta_D}(G_{\theta_G}(\Delta(I_n^{HR}))))) \tag{6}$$

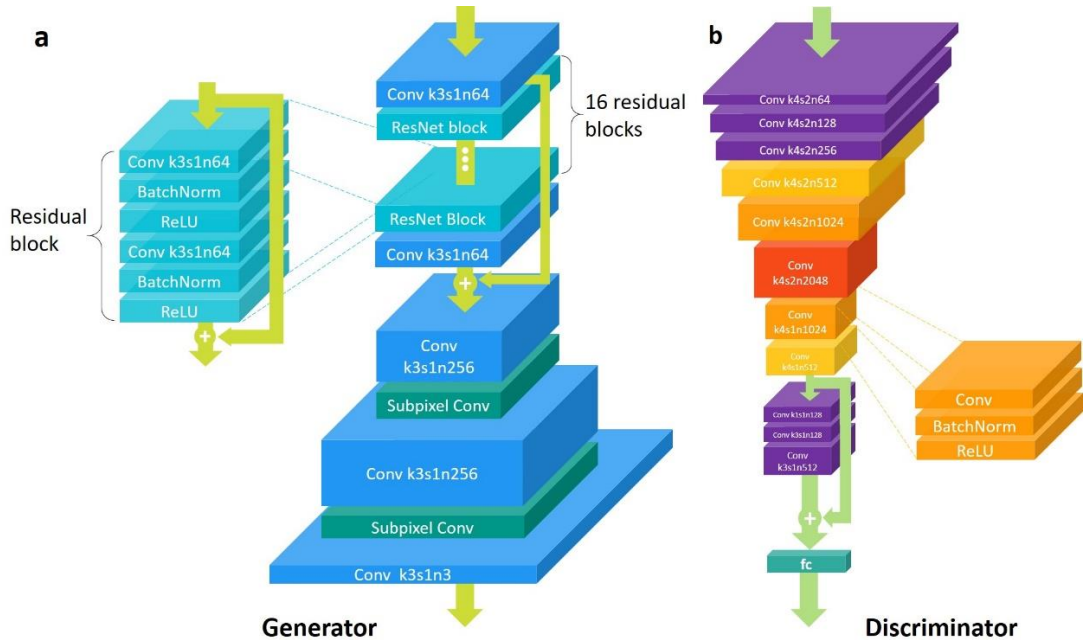

**Figure 4. The architecture of GAN. a,** Architecture of the generator. Conv and ResNet is the abbreviation of Convolutional layer and Residual network block. The parameters of the convolutional layers is given in the format "*k-s-n*", where *k* is the kernel size, *s* is the strides and *n* is the number of feature maps (i.e. the output channels of the layer). The depth of each convolutional layer roughly denotes the number of its feature maps, and the lateral dimensions denotes the size of its input. Totally, there are 16 residual blocks in the generator. **b,** Architecture of the discriminator. Each convolutional layer in the discriminator is the combination of a convolution layer, a batch normalization operation and a ReLU activation function.

## 2.4. Characterization of super-resolution GAN.

The capability of GAN is first characterized through imaging a negative USAF resolution target (Thorlabs R3L3S1N) with highest resolution of 228 line pairs per mm (lpm). We captured HR and corresponding LR images under a macro microscope (Olympus MVX10) with ×10 and ×2 total magnifications, respectively. Due to the simple pattern of the test target, an image registration was applied to match their corresponding FOV, forming strictly aligned HR and LR pairs for the GAN training. Considering the limited number of experimentally obtained samples, we applied a geometric transformation, such as translation and rotation, to these paired images to further expand the dataset. Finally, 1008 groups of HR and LR pairs were imported into the GAN network for training. Another large FOV, LR measurement was used to validate the converged network (Fig. 5b). As shown in Fig. 5c, d, the 5x-enhanced reconstructions have a significant improvement compared to the raw images. Due to the small magnification factor as well as limited numerical aperture, the raw ×2 image can hardly discern the high-frequency stripes in USAF target (Fig. 5b, cyan box for 114, and orange box for 228 lpm). The RFGANM reconstruction, in contrast, has resolved the finest part of USAF target (Fig. 5d2, 228 lpm). The GAN-reconstruction results are further compared with the realistic measurement under a x10 magnification (Fig. 5c3, d3), showing a good structural similarity (SSIM) to the high-resolution ground truth. The linecuts through the resolved line pairs (Fig. 5d) by each method are quantified in Fig. 5e, revealing a substantially improved resolution by GAN.

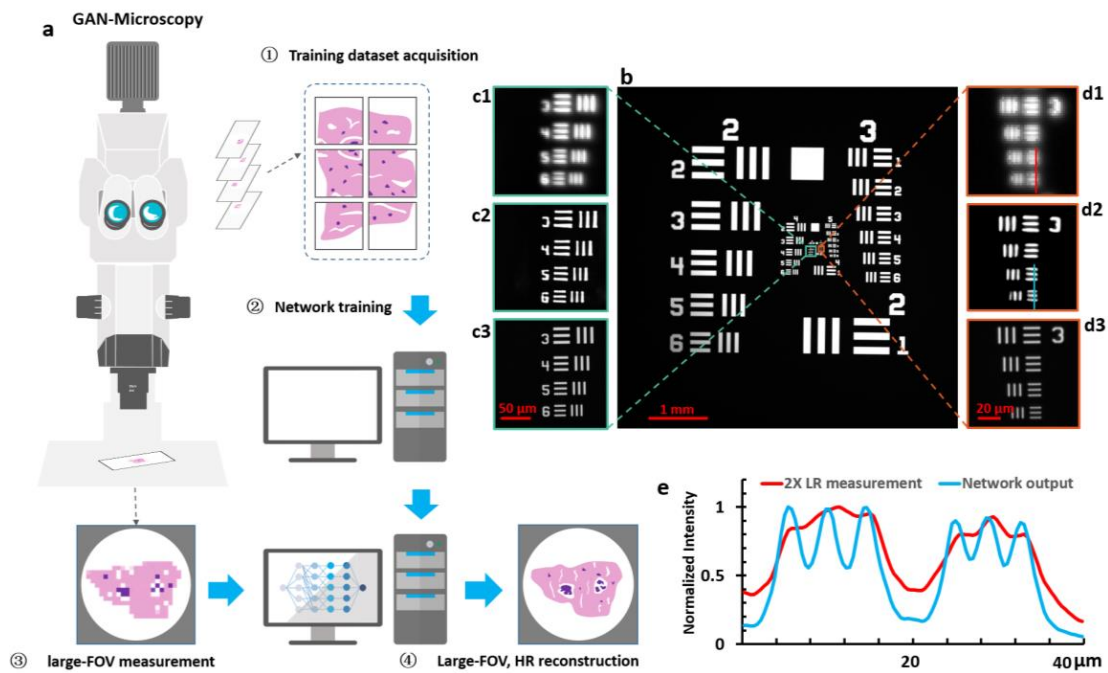

**Figure 5. Resolution characterization of GAN using USAF target**. a, the initial implementation of RFGANM, including HR training data acquisition, GAN network training, LR image acquisition of the sample, and HR reconstruction. **b,** Image of the USAF resolution target taken under ×2 low-magnification of a macro-zoom microscope. **c1** and **d1**, Magnified views of the raw image, with a pixel size of 2.13 μm. **c2** and **d2**, Reconstructions of a well-trained GAN-generator, with an enhancement factor of 5 (reconstructed pixel size, 0.425 μm). **c3** and **d3,** Corresponding high-resolution images taken under ×10 magnification. **e,** Intensity plot of linecuts (shown in d1 and d2) for each method, indicating that RFGANM prototype provides substantively improved contrast and resolution (FWHM ~1.7 μm) that potentially enable subcellular-level imaging across a centimeter large scale.

## 2.5. RFGANM of Dual-channel fluorescent BPAE cells.

We demonstrate our prototype by acquiring a wide-FOV dual-channel fluorescent image of a bovine pulmonary artery endothelial (BPAE) cell slide, as shown in Fig. 6a. The DAPI and Alexa Fluor 488 tagged nucleus and micro-filaments of the cells were imaged, trained and validated separately and finally merged into a pseudo-color display. To circumvent the high-demanding alignment between HR- and LR- measurements for their complicated subcellular patterns, we apply our image degrading model for LR training data generation instead of an image registration. Practically, we generate 4-times lower-resolution simulations using ×20 realistic images, to constitute the ×5 and ×20 training data pairs with their FOV intrinsically aligned. Vignette high-resolution views of the RFGANM outputs are provided in Fig. 6c-f with a reconstructed pixel size of 0.325 μm. The imaging FOV is around 24 mm$^2$, the same as that from a ×4 objective (Plan APO, 0.1 NA, Olympus), whereas the maximum achieved resolution is similar to that of a typical ×20 objective (Plan APO, 0.45 NA, Olympus). The conventional microscope images taken with ×20 and ×4 lenses are shown for comparison in Fig. 6c2 and d2, c3 and d3, respectively. Fig. 6c4 and d4 are the deconvolution results of c3 and d3, by Lucy-Richardson method, which provide no significant resolution promotions at all. It is noteworthy that beside the superior SISR capability, reconstructed image also shows a large depth of field (DOF) which is inherited from the ×4 source measurement. As a result, it behaves even better than the ×20 HR image in regions such as Fig. 6d2, where ×20 measurement is slightly out of focus due to its relatively small DOF. This underlying robustness of RFGANM indicates its easy implementation on a broad range of samples with natively unsmooth surfaces.

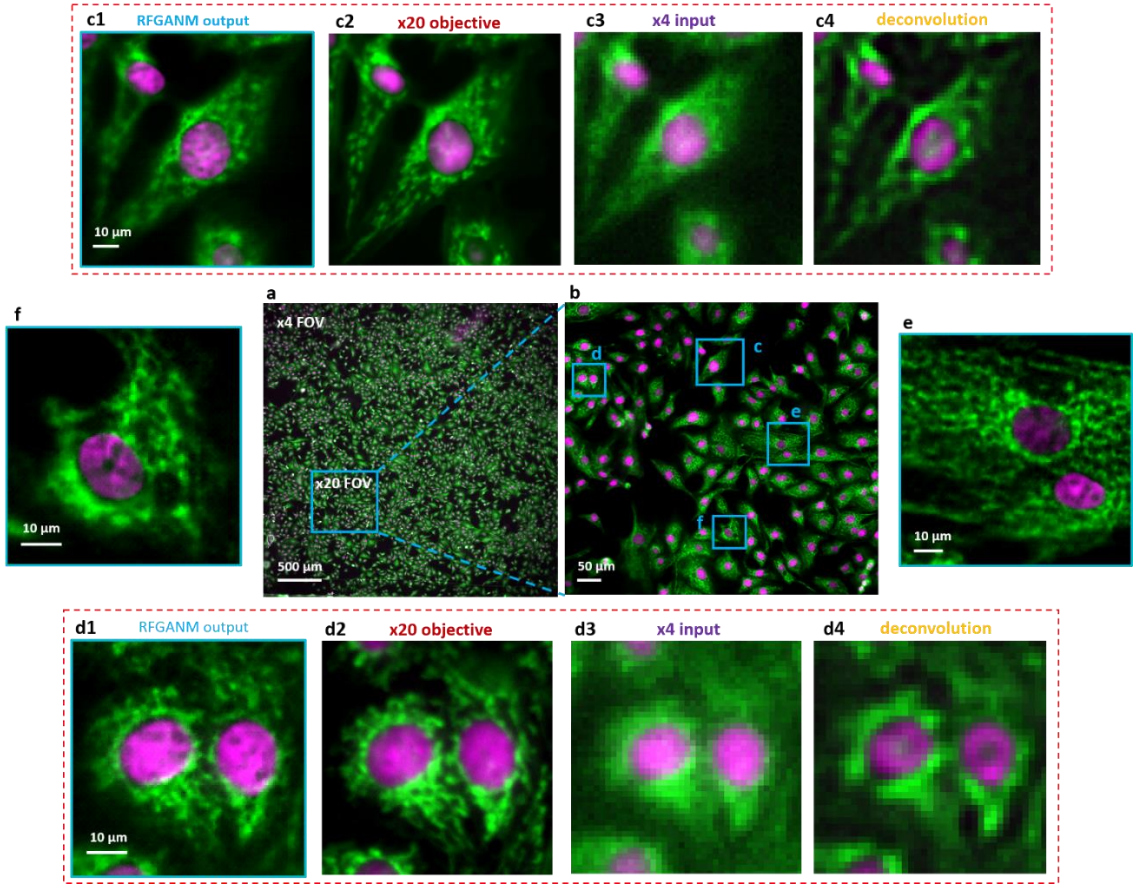

**Figure 6. Dual-color fluorescence imaging via RFGANM. a**, RFGANM-reconstruction of a wide-FOV fluorescence image of BPAE cells specifically labelled with DAPI and Alexa Fluor-488, at nucleus and skeletons, respectively. **a** and **b** show the imaging FOV of x4 (which RFGANM inherits) and x20 objectives (high-resolution conventional microscopy), respectively, by using a sCMOS camera (sensor area 1.33 x1.33 cm). **c1**, **d1**, **e** and **f**, high-resolution views of the selected regions (blue) in **a**. **c2** and **d2** are taken under a conventional wide-field fluorescence microscope with a x20/0.45 objective; **c3** and **d3** under x4/0.1 objective; and **c4**, **d4** are deconvolution results of **c3** and **d3** respectively.

## 2.6. High-throughput, gigapixel color imaging of pathological slides.

Several biomedical studies, such as histology and pathology, intrinsically need the observation of both large-scale statistics of tissue and local details of the cells. Here we further apply RFGANM to imaging of human tissue sections that are from a normal prostate and a prostate tumor. HR reference images were taken by a macro-zoom microscope (Olympus mvx10), using x10 magnification plus a color camera (QHY247C, 3.9 μm pitch size). The corresponding x2.5 LR training images were down-sampled from the HR measurements via the degrading model. The real x2.5 measurements of the tissue slides have a large FOV of about 64 mm$^2$, which almost covers the entire pathological sections (Fig. 7a,f). However, the magnified views show that the high-frequency details are largely decimated (Fig. 7c1, i1), and the deconvolution results by Lucy-Richardson method create artificial details that don't exist in ground truth. The x10 HR measurements contain abundant structure details, as shown in Fig. 7c2, i2, but suffer from much smaller FOV (~4 mm$^2$, inset in Fig. 7a). RFGANM, in contrast, is capable of providing numerous subcellular details across the entire sample scale. Vignette high-resolution views from different areas of the RFGANM reconstruction are shown in Fig. 7b-e and g-j, to verify its strong imaging capability. The RFGANM-reconstructed pathology slides encompass 0.4 gigapixels, with an effective SBP being ~0.1 gigapixels, which is 16-times higher than that of a conventional microscope. It is also noted that the pathological tissue slides often contain a lot of textures at various scales, thus posing a big challenge to the performance of super-resolution models. Our approach, under this circumstance, still achieves a high-fidelity result with high-frequency textures recovered.

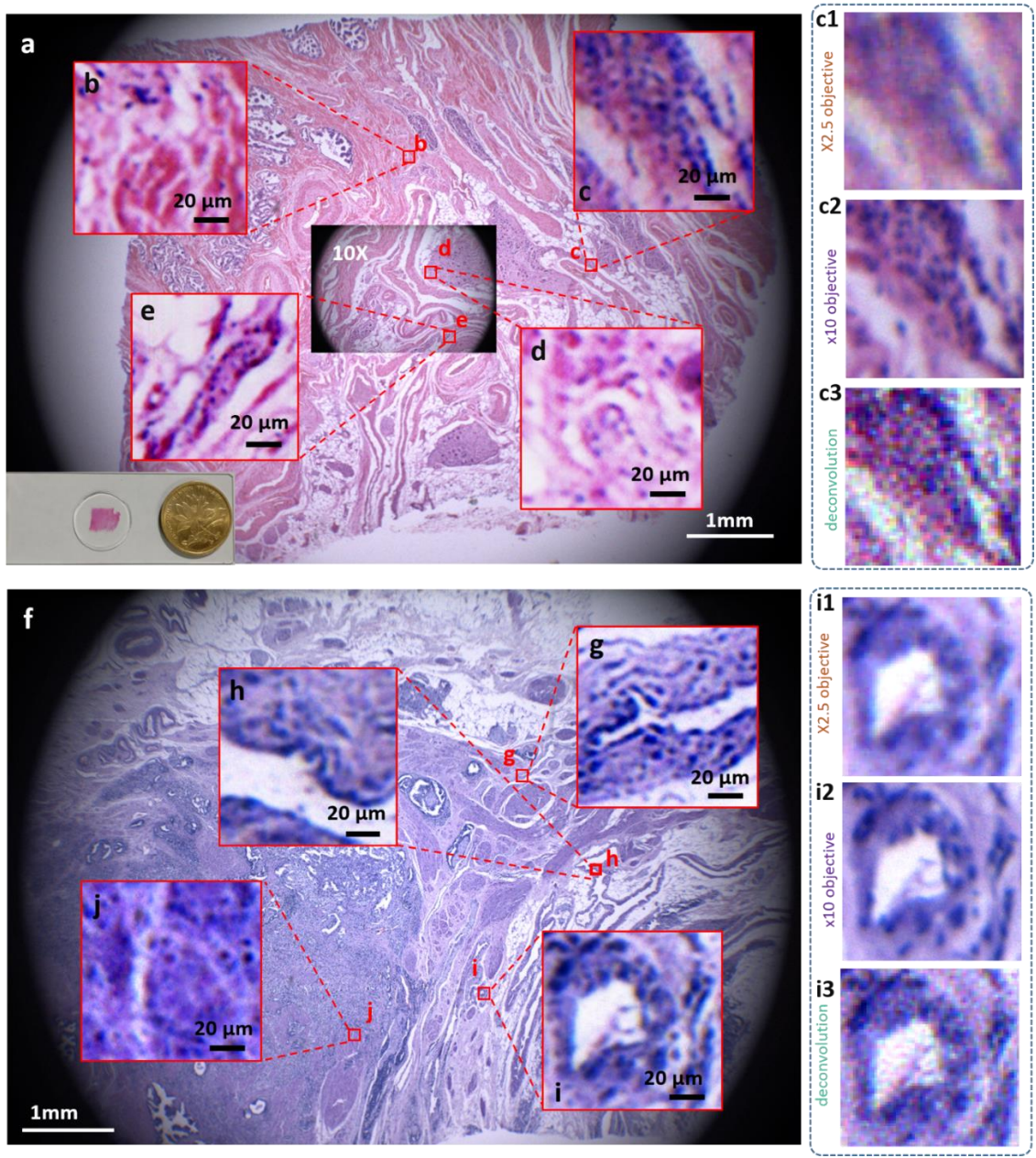

**Figure 7. Gigapixel color imaging of prostate tissue slides. a and f**, RFGANM color images of normal prostate histology slide and prostate cancer pathology slide, respectively. The achieved effective SBP here is ~0.1 gigapixels. **b-e**, **g-j** shows the vignette high-resolution views of the image in a and f. c1 and i1, the x2.5 input images of RFGANM to get c and i; c2 and i2, images taken by an Olympus macro-zoom microscope under x10 magnifications; c3 and i3, deconvolution results of x2.5 inputs c1 and i1, respectively, for comparison with RFGANM results.

*2.7. RFGANM in light-sheet imaging of selective planes in whole mouse brain.*

The advent of light-sheet fluorescence microscopy (LSFM) has greatly revolutionized three-dimensional (3-D) imaging of objects from cells to animals[36]. Compared to the epifluorescence methods, such as wide-field and confocal microscopes, LSFM has the advantages of tightly confined excitation, relatively low phototoxicity and high speed[37]. As a reference point, recent integration of LSFM and tissue-clearing has become an important alternative to conventional histology imaging approaches. However, even for current LSFM, the optical throughput of the system optics remains insufficient to *intoto* map the cellular information throughout a specimen of large volume size; for example, for visualization of the fine neuronal networks across a mouse brain. Tile imaging is the commonly-used approach to artificially increase the SBP, and realize high-resolution imaging of large specimens[38, 39]. Despite the compromised speed induced by repetitive mechanical stitching, the high illumination/ detection NA configuration in tile imaging greatly limits the fluorescence extraction from deep tissue of the thick specimens. We demonstrate, instead of commonly-used tile stitching, which significantly sacrifices the throughput and limit the signal extraction from deep tissue, the integration of RFGANM with light-sheet imaging can achieve high-resolution imaging of selective sectional planes in a whole adult mouse brain. We first constructed a macro-view light-sheet geometry with wide laser-sheet illumination and large-FOV detection (Fig. 8a), which can fully cover an optically-cleared P30 mouse brain (Tg: Thy1-GFP-M). 200 consecutively-illuminated transverse planes in the middle of the brain (depth 2 to 3 mm) were recorded (Fig. 8b), with their maximum-intensity-projection (MIPs) showing the global distribution of the neurons. The raw plane images simply accept the limited resolution from the macro-view LSFM system optics, hence the densely-packed neuronal fibers remain dim. The super-resolved image is then instantly obtained by RFGANM, with a reconstructed pixel size of 0.53 μm (Fig. 8c-h). The - result is furthermore compared to higher-magnification light-sheet measurements (×6.4 detection) to confirm the authenticity of the computation. In Fig. 8d3, the neuronal dendrites identified by each method reveal substantially improved resolution from RFGANM. Therefore, besides conventional epifluorescence methods, RFGANM is proven to be the same efficient to the LSFM imaging, which together are capable of rapidly obtaining the high-resolution signals from arbitrary planes of intact large tissues. Furthermore, in the light of the strong 3-D imaging capability of LSFM, RFGAN-LSFM is possibly to be extended to the third dimension in the future, to achieve high-throughput, high-resolution volumetric mapping of whole specimens, such us intact organs, and whole embryos.

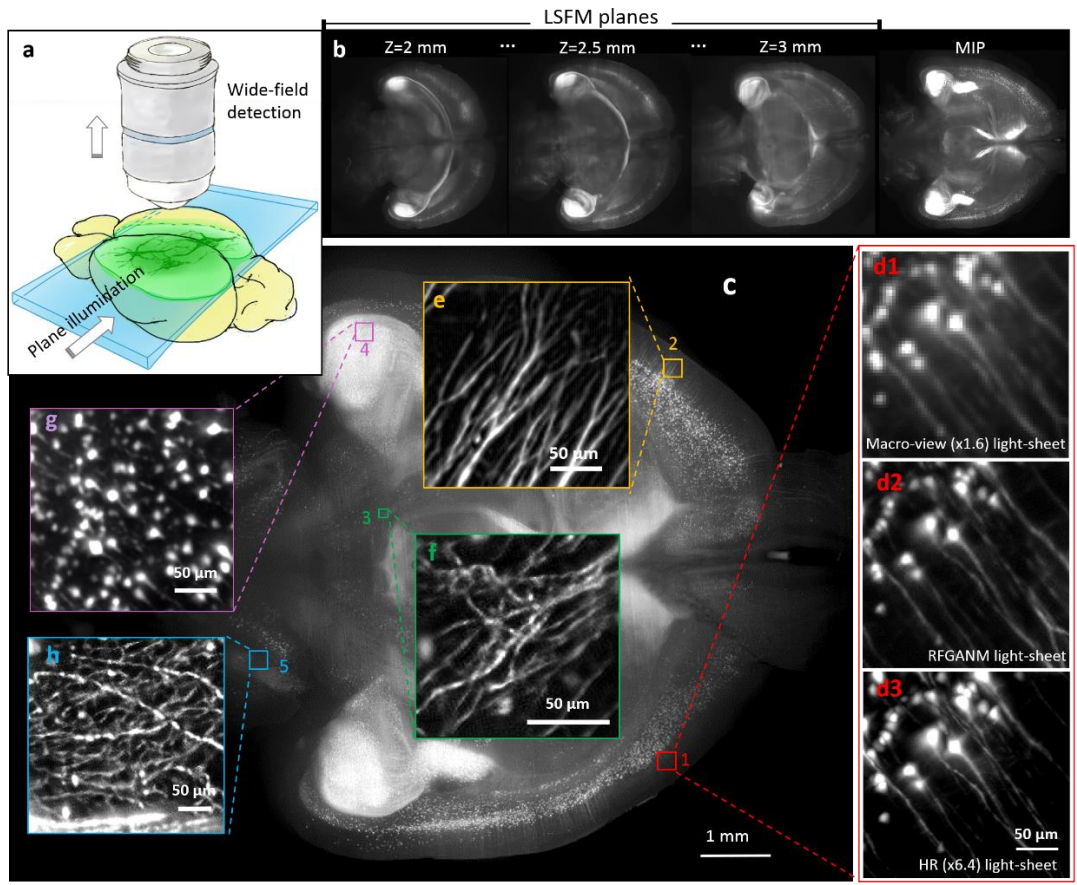

**Figure 8. GAN-based light-sheet fluorescence microscopy. a**, schematic of light-sheet imaging geometry for optical sectioning in the deep of intact organ. **b**, high-contrast plane images of the mouse brain. **c**, a 1mm-thick transverse slice of the whole brain, which is the MIP of 200 consecutively illuminated planes. The image was then super-resolved with 4 times enhancement, and compared to raw x1.6 (**d1**) and x6.4 measurements (**d3**). **e-h**, magnified reconstruction views of selective regions of cortex, telencephalon, hippocampus, and cerebellum.

*2.8. Quantitative evaluation of image similarity*

We calculate the peak signal-to-noise ratio (PSNR[dB]) and the structural similarity (SSIM)[40] index between the RFGANM outputs and the high-resolution measurements of the fluorescent BAEP cells (Table 1), to quantify the accuracy of the RFGANM. For comparison, PSNR and SSIM index between the bicubic interpolation followed by deconvolution of the low-resolution measurements and the corresponding high-resolution measurements are also computed. As a reference, the PSNR and SSIM of the high-resolution measurement to itself is ∞ and 1 respectively. Each test group contains 16 samples with a size of 384*384 pixels. We compute their PSNR and SSIM respectively and take the averaged value of all the 16 samples as final results. Apparently, SSIM and PSNR of the network reconstructions are both far better than that of the deconvolution results of bicubic interpolation of the LR measurements. The increase of PSNR from 19.63dB to 27.26dB validates the remarkable resolution enhancement, meanwhile the high-level SSIM index of 93.17% proves the authenticity of reconstruction.

Table 1. PSNR and SSIM between the GAN-reconstructed results and the realistic HR measurements

|  | SSIM | PSNR |
|---|---|---|
| Deconvolution of bicubic interpolation | 0.8093 | 19.6299 |
| Network reconstruction | 0.9317 | 27.2610 |
| High resolution Target | 1 | ∞ |

*2.9. The robustness of the network*

The training process of the GAN network is fundamentally a process in which the neural network extracts features from the training data and learns their roles in the corresponding LR and HR counterparts. Theoretically a network trained by one type of sample images should be also applicable to similar types of samples. For example, we can reasonably speculate that a GAN generator well-trained by healthy prostate data can work with the prostate cancer tissue as well. To test this underlying robustness, we blindly apply the network trained by healthy prostate tissue images to reconstruct a low-resolution image of prostate cancer tissue. Its outputs are compared with those from a real prostate-cancer-data-trained network, as shown in Fig. 9 below. Both networks recover highly similar structures with similar qualities presented, capable of resolving high-resolution cellular details, such as the nucleus and textures. It strongly suggests that GAN network could be highly robust, implying that RFGANM can go further with being applied to the reconstruction of a variety of samples merely with single type of data training.

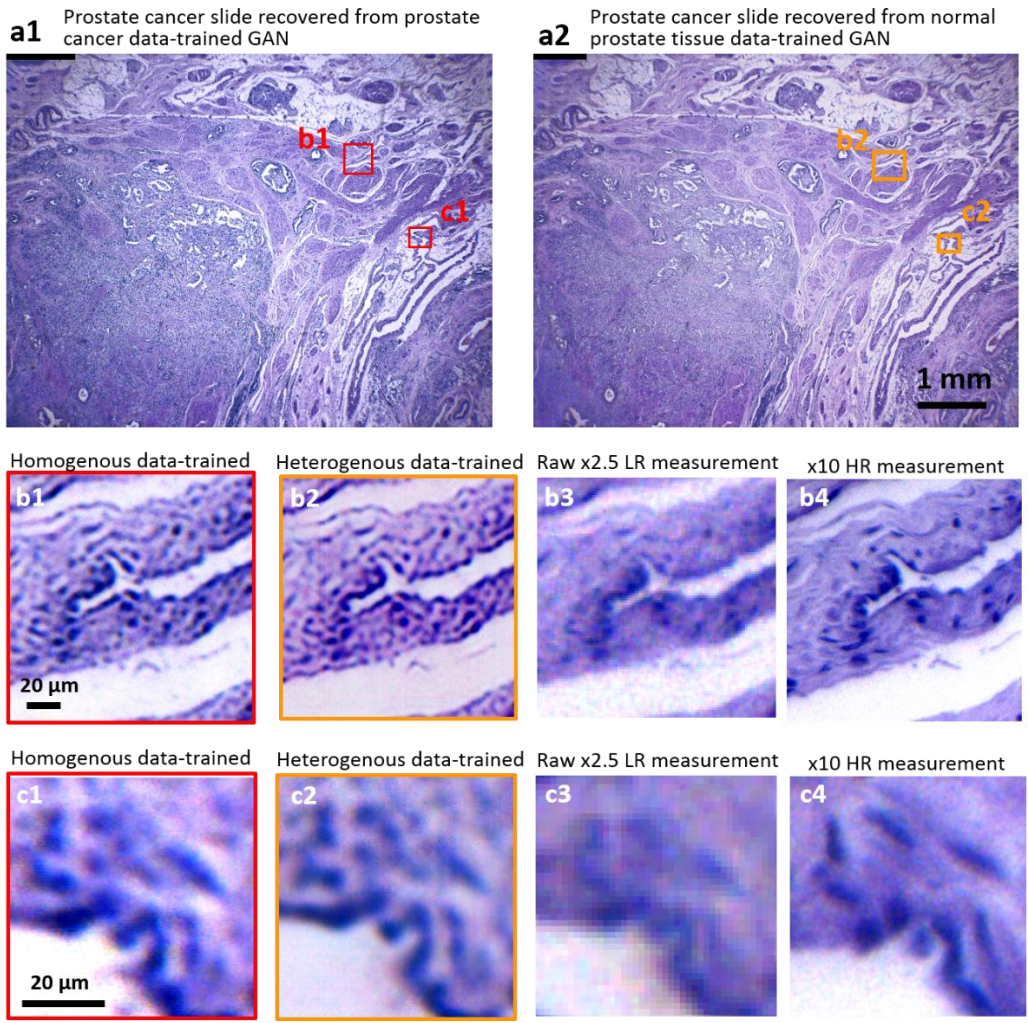

**Figure 9. Validating the robustness of RFGANM. a1** and **a2,** Wide-FOV reconstruction images of a prostate cancer slide using homogenous-data-trained (prostate cancer images) network and heterogenous-data-trained (healthy prostate images) network. **b1-c1**, and **b2-c2**, Vignette high-resolution views of the image in **a1** and **a2**, respectively. **b3-c3**, and **b4-c4**, Images taken by a macro-zoom microscope (mvx10) under a x2.5 and a x10 magnification, respectively, for comparison.

## 3. Methods

### 3.1. Sample preparation.

The fluorescent images in our experiments were taken on bovine pulmonary artery endothelial cells (BPAEC) that were fixed and multi-labeled before imaging. In cell preparation, after cell fixation and permeabilization, F-actin of cells was stained by Alexa Fluor® 488 phalloidin (Thermo Fisher Scientific Inc.), and the nuclei were counterstained using the blue-fluorescent DNA stain DAPI. The pathological tissue slides for bright-field microscopy were healthy human prostate tissue and human prostate cancer tissue, respectively, stained with hematoxylin-eosin after paraffin-embedding and automatic slicing. For light sheet imaging, an uDISCO cleared P30 mouse brain is used.

*3.2. Training implementations.*

Our model is implemented based on Google's deep learning framework, TensorFlow (version r1.3), and trained on an Inspur® server with two NVidia Tesla P100 GPUs. Initiating with a batch size of 16 and a fixed learning rate of $10^{-4}$, we trained the network for 200 epochs, which took about 48 hours.

*3.3. Inference process.*

In the inference phase after network training, the experimentally captured LR images for validation are cut into a bunch of small pieces with overlaps with each other, and then input into the network for super-resolved reconstruction piece by piece. Afterwards, all these output pieces are stitched into one whole image that possesses both large FOV and high resolution. The stitching process is achieved by matching the overlapped regions, which is very robust and accurate. The inference process is quite fast. An image piece of 100x100 pixels size takes less than 0.01 second to be super-resolved into a 400x400 pixels image, even on an ordinary Windows laptop with Intel Core i5 CPU.

*3.4. Imaging Setups.*

There are several kinds of microscopy images in our experiment: the bright-field grayscale resolution test target images, the dual-color fluorescent BPAE images, the bright-field color images of two types of tissue slides, and the light sheet images of mouse brain. Images of resolution target were recorded by a Photometrics IRIS15 camera (pitch size 4.25 μm), with the HR and LR images taken under x10 and x2 magnifications of an Olympus MVX10 microscope, respectively. The BPAE cells were imaged under an Olympus IX73 microscope equipped with a HAMAMATSU ORCA-Flash 4.0-V2 camera (pitch size 4.25 μm). In both fluorescent channels, the HR training images and LR validation images were taken under a x20/0.45 and x4/0.1 objective, respectively. For pathology slide imaging, a QHY247C color camera (pitch size is 3.9 μm) was used on the Olympus MVX 10 microscope to capture the healthy prostate/prostate cancer tissues stained with hematoxylin-eosin. The HR training and LR validation images were then taken under x10 and x2.5 magnifications, respectively. The sectional images of intact mouse brain were obtained by a macro-view light-sheet system, which comprised a self-made dual-side laser-sheet illumination and large-FOV wide-field detection by Olympus MVX10 microscope body. The images were sequentially recorded using Photometrics IRIS15 camera under ×1.6 and 6.4 magnifications.

## 4. Conclusion

We have demonstrated a deep learning-based microscopy method without the requirement of extra registration procedure in training course, which can significantly improve the resolution of conventional wide-field and cutting-edge light-sheet fluorescence microscopes, and greatly increase the imaging throughput for whole biomedical specimens. We apply a state-of-the-art GAN network to deeply learn how to map from the low-resolution microscopy images to their high-resolution counterparts. For cell and tissue images that contain complicated patterns, their low-resolution training data are artificially generated and intrinsically registered to the high-resolution training images via a degradation model. This step has simplified the data preprocessing and improved the robustness of the GAN network. Once the model training being accomplished, the well-established AI agent is capable of quickly reconstructing a large FOV, super-resolution image of new sample based on a single low-resolution snapshot taken by an ordinary optical microscope. Besides the improved resolution that has been verified by imaging of resolution target and PSNR analysis, the structure similarity to the sample ground truth has also been quantified, at a high level of over 90%. We also prove that this RFGANM method is very robust, readily applicable to most forms of microscopy data such as bright-field images, epifluorescence images, and light-sheet fluorescence images. It significantly extends the SBP of these microscope systems neither at the cost of acquiring multiple frames nor relying on the retrofit of conventional microscope system. Therefore, RFGANM has a high temporal performance similar to previous example-based SISR methods, but shows a much better image quality that is comparable to those multi-frame SR methods. As a reference point, it produces a 0.38 gigapixel digital pathology slide at 1 μm resolution, with an acquisition time of 0.01 second and computation time of less than 1 second (with normal CPU desktop). This high-resolution combined with high-throughput capability renders RFGANM a valuable tool for many applications, such as tissue pathology and neuroanatomy. Furthermore, though currently we demonstrate the combination of deep learning and convolutional neural network with optical microscopy in form of 2-D imaging of *exvivo* samples, we can reasonably expect that provided its superior spatial-temporal performance, this methodology will be also applicable to both 3-D microscopy and highly dynamic process.

**Funding**

1000 Youth Talents Plan of China (P.F.), National Key R&D program of China (P.F., 2017YFA0700500), Research Program of Shenzhen (P.F., JCYJ20160429182424047).

**Acknowledgements**

The authors acknowledge the selfless sharing of the GAN source codes from Hao Dong (hao.dong11@imperial.ac.uk), as well as the contributions of Tinting Zhu for assistance with fluorescent sample preparation. The authors thank Wenbin Jiang and Yang Ma for their assistance with GPU-based computation.

**Disclosures**

The authors declare that there are no conflicts of interest related to this article.

**References**

1. Zheng, G., R. Horstmeyer, and C. Yang, *Wide-field, high-resolution Fourier ptychographic microscopy.* Nature Photonics, 2013. **7**(9): p. 739.
2. Lohmann, A.W., et al., *Space–bandwidth product of optical signals and systems.* Journal of the Optical Society of America A, 1996. **13**(3): p. 470-473.
3. Hilgen, G., et al. *High Resolution Large-Scale Recordings of Light Responses from Ganglion Cells in the Developing Mouse Retina*. in *FASEB Conference on Retinal Neurobiology and Visual Processing*. 2014.

4. Imfeld, K., et al., *Large-scale, high-resolution data acquisition system for extracellular recording of electrophysiological activity.* IEEE transactions on bio-medical engineering, 2008. **55**(8): p. 2064-73.
5. Imfeld, K., et al., *Large-Scale, High-Resolution Data Acquisition System for Extracellular Recording of Electrophysiological Activity.* IEEE Transactions on Biomedical Engineering, 2008. **55**(8): p. 2064.
6. Brown, M. and D.G. Lowe, *Automatic Panoramic Image Stitching using Invariant Features.* International Journal of Computer Vision, 2007. **74**(1): p. 59-73.
7. Hillman, T.R., et al., *High-resolution, wide-field object reconstruction with synthetic aperture Fourier holographic optical microscopy.* Optics Express, 2009. **17**(10): p. 7873.
8. M, K., et al., *High-speed synthetic aperture microscopy for live cell imaging.* Optics letters, 2011. **36**(2): p. 148.
9. Gutzler, T., et al., *Coherent aperture-synthesis, wide-field, high-resolution holographic microscopy of biological tissue.* Optics Letters, 2010. **35**(8): p. 1136.
10. Luo, W., et al., *Synthetic aperture-based on-chip microscopy.* Light Science & Applications, 2015. **4**(3).
11. Zheng, G., et al., *Sub-pixel resolving optofluidic microscope for on-chip cell imaging.* Lab on A Chip, 2010. **10**(22): p. 3125-3129.
12. Zheng, G., et al., *The ePetri dish, an on-chip cell imaging platform based on subpixel perspective sweeping microscopy (SPSM).* Proceedings of the National Academy of Sciences of the United States of America, 2011. **108**(41): p. 16889.
13. Luo, W., et al., *Pixel super-resolution using wavelength scanning.* Light Science & Applications, 2016. **5**(4): p. e16060.
14. Xu, W., et al., *Digital in-line holography for biological applications.* Proceedings of the National Academy of Sciences of the United States of America, 2001. **98**(20): p. 11301-5.
15. Denis, L., et al., *Inline hologram reconstruction with sparsity constraints.* Optics Letters, 2009. **34**(22): p. 3475.
16. Greenbaum, A., et al., *Increased space-bandwidth product in pixel super-resolved lensfree on-chip microscopy.* Scientific Reports, 2013. **3**(3): p. 1717.
17. Elad, M. and Y. Hel-Or, *A fast super-resolution reconstruction algorithm for pure translational motion and common space-invariant blur.* IEEE transactions on image processing : a publication of the IEEE Signal Processing Society, 2001. **10**(8): p. 1187.
18. Farsiu, S., et al., *Fast and robust multiframe super resolution.* IEEE Transactions on Image Processing, 2004. **13**(10): p. 1327-1344.
19. Vandewalle, P., S. Sü, and M. Vetterli, *A Frequency Domain Approach to Registration of Aliased Images with Application to Super-Resolution.* Eurasip Journal on Advances in Signal Processing, 2006. **2006**(1): p. 1-14.
20. Gustafsson, M.G.L., *Surpassing the lateral resolution limit by a factor of two using structured illumination microscopy.* Journal of Microscopy, 2010. **198**(2): p. 82-87.
21. MG, G., *Nonlinear structured-illumination microscopy: wide-field fluorescence imaging with theoretically unlimited resolution.* Proceedings of the National Academy of Sciences of the United States of America, 2005. **102**(37): p. 13081.

22. Yang, J., et al. *Image super-resolution as sparse representation of raw image patches*. in *Computer Vision and Pattern Recognition, 2008. CVPR 2008. IEEE Conference on*. 2008.
23. Timofte, R., V.D. Smet, and L.V. Gool, *A+: Adjusted Anchored Neighborhood Regression for Fast Super-Resolution*. 2014: Springer International Publishing. 111-126.
24. Ozcan, A., et al., *Deep learning microscopy.* 2017.
25. Rivenson, Y., et al., *Deep learning enhanced mobile-phone microscopy.* Acs Photonics, 2017.
26. Ouyang, W., et al., *Deep learning massively accelerates super-resolution localization microscopy.* Nature Biotechnology, 2018. **36**(5).
27. Nehme, E., et al., *Deep-STORM: super-resolution single-molecule microscopy by deep learning.* 2018.
28. Recht, N.B.E.J.H.B.B., *DeepLoco: Fast 3D Localization Microscopy Using Neural Networks.* BioRxiv, 2018. **236463**.
29. Hongda Wang, Y.R., Yiyin Jin, Zhensong Wei, Ronald Gao, Harun and L.A.B. Günaydın, Aydogan Ozcan, *Deep learning achieves super-resolution in fluorescence microscopy.* BioRxiv, 2018. **309641**.
30. Zhang, H., et al., *High-throughput, high-resolution Generated Adversarial Network Microscopy.* 2018.
31. Goodfellow, I.J., et al. *Generative adversarial nets*. in *International Conference on Neural Information Processing Systems*. 2014.
32. Shi, W., et al., *Real-Time Single Image and Video Super-Resolution Using an Efficient Sub-Pixel Convolutional Neural Network.* 2016: p. 1874-1883.
33. Ledig, C., et al., *Photo-Realistic Single Image Super-Resolution Using a Generative Adversarial Network.* 2016.
34. Johnson, J., A. Alahi, and L. Fei-Fei, *Perceptual Losses for Real-Time Style Transfer and Super-Resolution.* 2016: p. 694-711.
35. Simonyan, K. and A. Zisserman, *Very Deep Convolutional Networks for Large-Scale Image Recognition.* Computer Science, 2014.
36. Huisken, J., et al. *EH: Optical sectioning deep inside live embryos by selective plane illumination microscopy*. in *Science*. 2004.
37. Keller, P.J., et al., *Reconstruction of Zebrafish Early Embryonic Development by Scanned Light Sheet Microscopy.* Science, 2008. **322**(5904): p. 1065-1069.
38. Becker, K., et al., *Ultramicroscopy: 3D reconstruction of large microscopical specimens.* Journal of Biophotonics, 2008. **1**(1): p. 36-42.
39. Guan, Z., et al., *Compact plane illumination plugin device to enable light sheet fluorescence imaging of multi-cellular organisms on an inverted wide-field microscope.* Biomedical Optics Express, 2015. **7**(1): p. 194.
40. Wang Z., B.A.C., Sheikh H. R., Simoncelli E. P. , *Image Quality Assessment: From Error Visibility to Structural Similarity.* IEEE Trans. Image Process, 2004. **13**: p. 13.